\documentclass[12pt]{article}
\usepackage{authblk}
\usepackage{amsmath}
\usepackage{amssymb}
\usepackage[left=20mm, top=30mm, right=20mm, bottom=30mm]{geometry}
\usepackage{graphicx}
\usepackage{subfig}
\usepackage[font=scriptsize]{caption}
\begin{document}
\title{Three-dimensional slowly rotating black hole in Einstein-power-Maxwell theory}
\author{M. B. Tataryn\footnote{E-mail: misha.physics@gmail.com}, M. M. Stetsko\footnote{E-mail: mstetsko@gmail.com}}
\affil{\small Department for Theoretical Physics,\\Ivan Franko National University of Lviv,\\12 Drahomanov Str., Lviv, 79005, Ukraine}
\date{}
\maketitle
\abstract{A three-dimensional slowly rotating black hole solution in the presence of negative cosmological constant in the Einstein-power-Maxwell theory is studied. It is shown that in the small rotation limit the electric field, diagonal metric function and thermodynamic properties in the small rotation limit are the same as for static case, whereas the small rotation gives in addition a non-diagonal metric function and magnetic field which are also small. For these functions cased by rotation of black hole it was obtained exact integral solution and analytic asymptotic solution.

}
\section{Introduction}
Black holes play an exceptional role in the nowadays physics study. Being one of the crucial objects of General Relativity, namely one of the present fundamental physical theories, they admit the research on extremal investigative area, and thus apparently make possible the improvement of our nature understanding not only regarding classical theory of gravity but also quantum gravity, string theory, cosmology, electrodynamics, thermodynamics, and thereby becoming a great theoretical laboratory for physics studies.

Particularly, the developed approach of black hole thermodynamics \cite{bek73}, permitting the thermodynamics describing of black holes in terms more customary to us inclines to assume the black hole physics along with other physical systems using the same notions. Another example is the applying of the Maxwell electrodynamics to the black hole systems as it firstly took place in the Reissner-Nordstr{\"o}m solution \cite{reis16}, which presents the four-dimensional static charged black hole.

Besides the well known four-dimensional black holes in Einstein gravity and Maxwell electrodynamics, there is a number of reasons to consider various generalizations and modifications of these theories applied to the black hole systems.

One of such possible generalizations is consideration of various spacetime dimensions. Such approach is widely used in modern physics and currently caused mainly by the studying of supergravity, string theory, particle physics in higher dimensions. The lower dimensional gravity is also studied, here it is appropriate to refer to the original BTZ black hole \cite{ban92,mar00}, where it was introduced a three-dimensional black hole solution and shown that it possesses the key features as its four-dimensional analogues. (2+1)-dimensional black holes were considered in several contexts, for example, black hole with power Maxwell source in scale-dependent theory \cite{rin17,rin18} and black hole with dilaton field in Einstein-power-Maxwell theory \cite{hendi17}.

Among generalizations and modifications of Einstein gravity it can be mentioned such research directions as, so-called, $f(R)$ gravity and scalar-tensor gravity, for more detail see, for example, \cite{hen_pan_saf_14} and \cite{stef07}, respectively.

Modifications of the Maxwell electrodynamics have long been known in physics, originally the Born-Infeld electrodynamics \cite{born34} and later the logarithmic and exponential electrodynamics which belong to the, so-called, Born-Infeld type nonlinear electrodynamics, see, for example, \cite{hen_all_14,shey14,hen16,tat19}. Besides this, the one of the natural possibility to generalize the Maxwell theory and which has been attracting a lot of attention nowadays is the Einstein-power-Maxwell theory in which the electromagnetic lagrangian enters as a power of the Maxwell invariant, namely $(F_{\mu\nu}F^{\mu\nu})^s$ with power $s$. One of the main prerequisites to consider this type of nonlinearity is due to the conformal invariance which is present only in the four-dimensional Maxwell theory. In the conformal invariant theory the electromagnetic stress-energy tensor is traceless, the scalar curvature vanishes and one can obtain the inverse square law for electric field, namely $\sim1/r^2$. It was shown (see, for example, \cite{hen10}) that the choice the power $s=d/4$ in $d$-dimensional spacetime leads to the conformally invariant charged black hole solutions. The studies in the context of the power Maxwell source can be found, for example, in \cite{mazh14} where a special values $s=1/2$ was examined. In \cite{has07} it was obtained the higher-dimensional static black hole solutions with the conformally invariant Maxwell source, whereas in \cite{has08} it can be found a static black hole solutions in $d\geqslant3$ spacetime dimensions for various values of nonlinear parameter $s$. In that paper it was also shown that the choice $d=2s+1$ leads to $\sim1/r$ dependence for electric field and as a consequence to the appearance of the logarithmic term in the metric similarly as taken place for linear Maxwell theory ($s=1$) in three dimensions (BTZ black hole). A slowly rotating black hole with spherical horizon structure in higher-dimensional ($d\geqslant4$) Einstein-power-Maxwell theory was studied in \cite{hen10}, whereas a slowly rotating toroidal black hole with flat horizon structure in four dimensions was examined by the authors of article \cite{pan19}. In this cases it was shown that thermodynamic properties of black holes in the small rotation limit are the same as in corresponding static cases. Rotating aspects of black holes were also recently investigated, namely it was constructed a (3+1)-dimensional rotating black hole in scale-dependent gravity \cite{cont20} and (3+1)-dimensional rotating polytropic black hole \cite{cont19}. Other possibilities of nonlinear electrodynamics can be found, for example, in \cite{gaet17,krugl15}.

Black hole thermodynamics in the Einstein-power-Maxwell theory was considered by many authors, see, for example, \cite{pan19,gon09}, whereas an extended phase space thermodynamics where cosmological constant is associated with thermodynamic pressure was reviewed, for example, in \cite{kub12,hen13}.

Besides, recently it was investigated spectrum of quasinormal modes of black holes in 2+1 dimensions with power Maxwell field in scale-dependent theory \cite{rincon18}, in (3+1)- \cite{panot18} and (4+1)- \cite{panotop20} dimensional Einstein-power-Maxwell theory.

As it was mentioned above there are many works on various-dimensional black holes charged with power Maxwell source in static and rotation cases and as far as we know there is not considered a three-dimensional black hole solution with power Maxwell field in case of rotation including even the small rotation limit. Therefore, in the present paper we examine a slowly rotating black hole solution in the presence of negative cosmological constant in three-dimensional Einstein-power-Maxwell theory and then consider thermodynamic behavior of this solution in the extended phase space thermodynamics. Also, in this way we naturally generalize our previous research \cite{tat19}, where we considered only static solution in three-dimensional Einstein gravity, but apart of the power-Maxwell case we also examined Born-Infeld, logarithmic and exponential electromagnetic field lagrangians. So, the structure of the present study is as follows: in Section \ref{solution} we write field equations and obtain the slowly rotating black hole solution. In Section \ref{charges} we calculate black hole mass and angular momentum, in Section \ref{thermodynamics} we consider thermodynamics of the black hole. Finally, in Section \ref{conclusions} we give some concluding remarks.
\section{Black hole solution}\label{solution}
We are going to obtain a three-dimensional slowly rotating black hole solution in Einstein-power-Maxwell theory in the presence of a negative cosmological constant (the case of anti-de Sitter space). Thereby we consider the rotating charged black hole and thus it is described by the full set of black hole macroscopic parameters, that is the black hole mass $M$, the total electric charge $Q$ and the angular momentum $J$. The bulk action in the three-dimensional Einstein-power-Maxwell theory with scalar curvature $R$, cosmological constant $\Lambda$ and the electromagnetic Lagrangian of the power Maxwell field $L=(-F)^s$ can be written in the form (taking $G\equiv1$).
\begin{equation}\label{act}
I[g_{\mu\nu},A_\mu]=\frac{1}{16\pi}\int d^3x\sqrt{-g}[R-2\Lambda+(-F)^s],
\end{equation}
where $g_{\mu\nu}$ is the metric tensor, $A_\mu$ is the electromagnetic potential, $g$ is determinant of the metric tensor $g_{\mu\nu}$ and $F=F_{\mu\nu}F^{\mu\nu}$ is the Maxwell invariant. The special value of nonlinearity parameter $s=1$ leads to linear Maxwell electrodynamics. Note that here we take a unit prefactor for the electromagnetic Lagrangian. Using the principle of least action, after variation (\ref{act}) with respect to $g_{\mu\nu}$ and $A_\mu$ one can obtain the equations for gravitational and electromagnetic fields respectively, called the Einstein-Maxwell equations
\begin{equation}\label{grav}
R_{\mu\nu}-\frac{1}{2}g_{\mu\nu}R+g_{\mu\nu}\Lambda=T_{\mu\nu},
\end{equation}
where $R_{\mu\nu}$ is the Ricci tensor and $T_{\mu\nu}$ is the stress-energy tensor
\begin{equation}\label{stress}
T_{\mu\nu}=\frac{1}{2}g_{\mu\nu}L-2\frac{\partial L}{\partial F}F_{\mu\beta}F_\nu^{\ \beta},
\end{equation}
\begin{equation}\label{em}
\partial_\mu\left(\sqrt{-g}\frac{\partial L}{\partial F}F^{\mu\nu}\right)=0.
\end{equation}

\begin{figure}[h]
\centering
\subfloat{\includegraphics[width=0.35\textwidth]{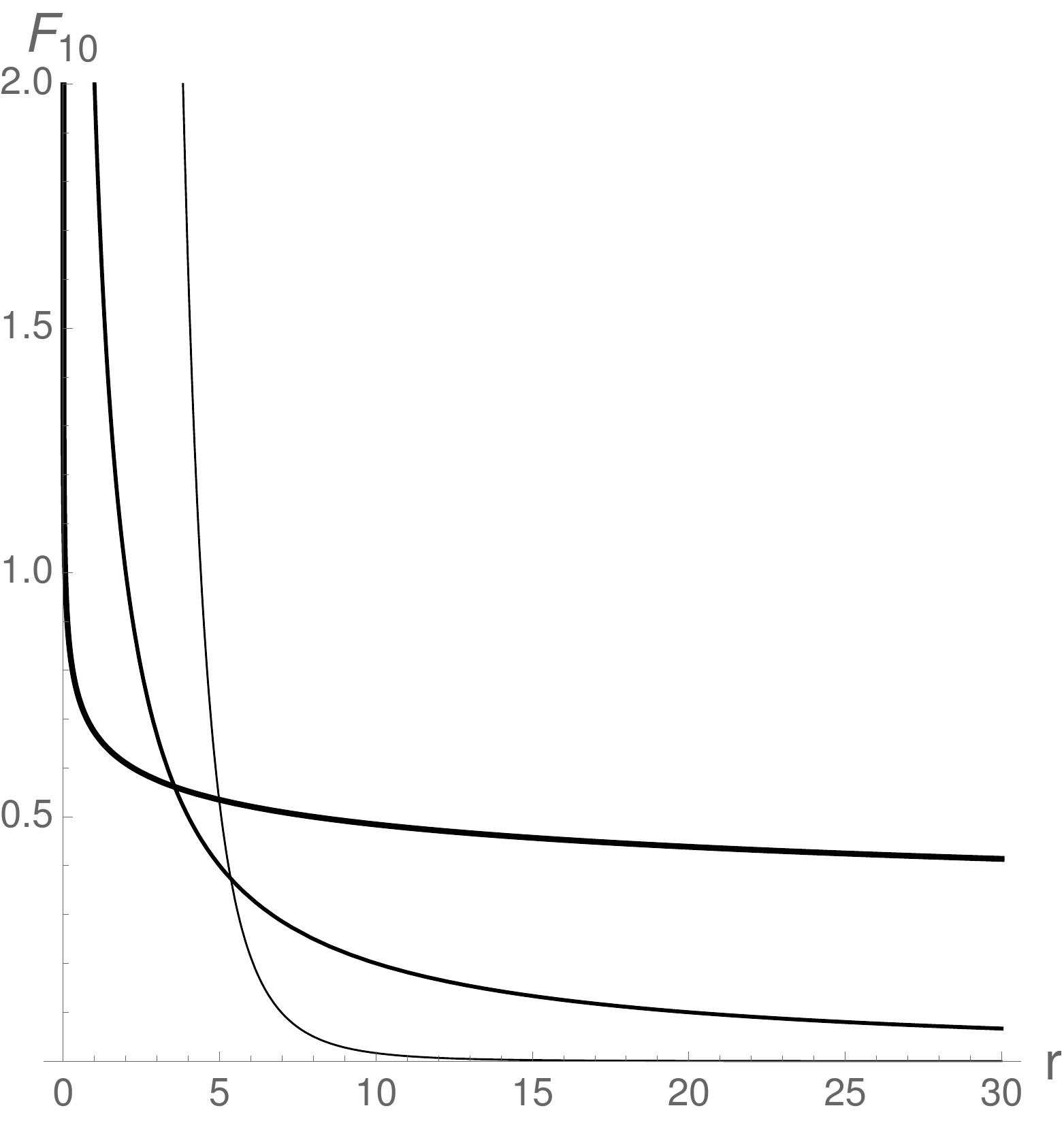}}
\qquad
\subfloat{\includegraphics[width=0.35\textwidth]{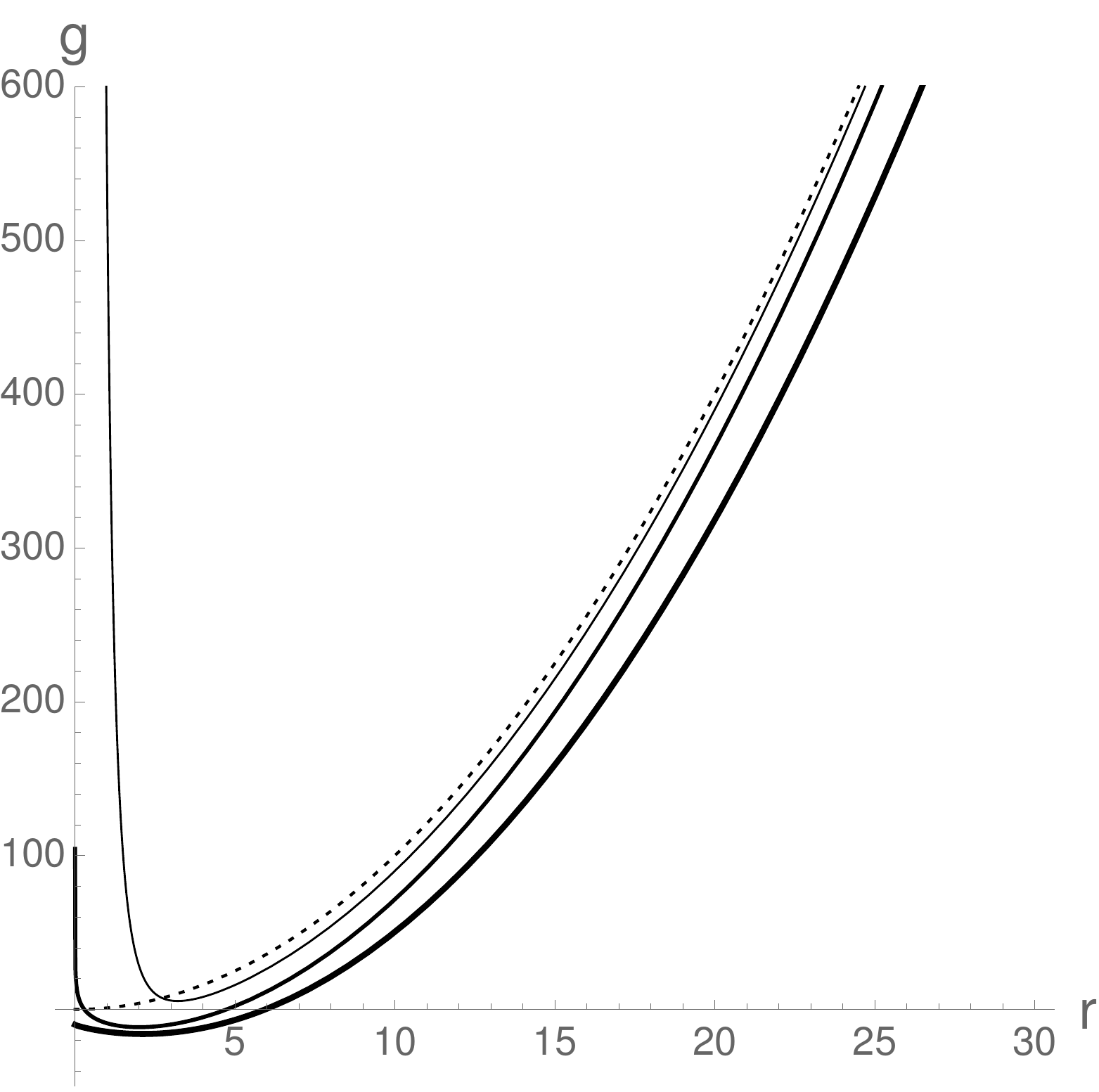}}
\caption{Electric field $F_{10}(r)$ (left) and diagonal metric function $g(r)$ (right) for $s=0.6$ (thin), $s=1$ (normal) and $s=4$ (bold) for parameters $q_1=2$, $m=10$, $\Lambda=-1$. Dots show the behavior of cosmological term $-\Lambda r^2$ of function $g(r)$.}
\label{fig_1}
\end{figure}

Since we study the three-dimensional stationary slowly rotating black hole solution the line element in spacetime with coordinates $x^0=t\in(-\infty,+\infty)$, $x^1=r\in[0,+\infty)$, $x^2=\varphi\in[0,2\pi)$ can be written in the form
\begin{equation}\label{met}
ds^2=-g(r)dt^2+\frac{dr^2}{g(r)}+r^2d\varphi^2+2ar^2f(r)dtd\varphi,
\end{equation}
where $g(r)$ and $f(r)$ are the metric functions, thereby we have dependences only on radial coordinate $r$, $a$ is the parameter of slow rotation, and this slow rotation means that  here and in the following we take into account only the terms linear over $a$. The setting $a=0$ in (\ref{met}) recovers the metric of the static black hole, thereby the parameter $a$ is responsible for rotation (slow) of black hole and is related to the angular momentum $J$ \cite{hen10}. For the chosen metric we have only radial electric $F_{10}$ and magnetic $F_{12}$ components of electromagnetic tensor, we also assume that $F_{12}\sim a$, so the Maxwell invariant $F=-2F_{10}^2$ is the same as for static case. The diagonal components of Eqs.~(\ref{grav})-(\ref{stress}) and the ``0''-component of Eq.~(\ref{em}) are the same as for the static case
\begin{equation}\label{0011}
g'+2\Lambda r+2^s(2s-1)rF_{10}^{2s}=0,
\end{equation}
\begin{equation}
g''+2\Lambda-2^sF_{10}^{2s}=0,
\end{equation}
\begin{equation}\label{0}
(2^{s-1}srF_{10}^{2s-1})'=0,
\end{equation}
where prime denotes derivative with respect to $r$. The system of equations (\ref{0011})-(\ref{0}) gives exact solutions for functions $F_{10}$ and $g$ which were written in \cite{tat19}\footnote{In that solutions we had different integration constants for different $s$, thus that formulas differ a bit from the ones given here.}
\begin{equation}\label{F10}
F_{10}=\left(\frac{q_1}{2^{s-1}sr}\right)^\frac{1}{2s-1},
\end{equation}
\begin{equation}\label{g1}
g=-\Lambda r^2-2q_1^2\ln r-m,\qquad s=1,
\end{equation}
\begin{equation}\label{gs}
g=-\Lambda r^2-\frac{2^{s-1}(2s-1)^2}{s-1}\left(\frac{q_1}{2^{s-1}s}\right)^\frac{2s}{2s-1}r^\frac{2(s-1)}{2s-1}-m,\qquad s\neq1,
\end{equation}
where $q_1$ and $m$ are integration constants related to the charge and mass of black hole, respectively. Here and in the following we consider that the argument $r$ in the logarithmic function is dimensionless. Imposing condition that the electric field $F_{10}$ must decrease monotonously with increasing $r$ and considering that $s$ takes values of the order of unity we obtain from Eq.~(\ref{F10}) the lower bound on the power $s$, namely $s>1/2$ since for $s<1/2$ we have the divergent behavior of $F_{10}$ when $r\to+\infty$ what is unphysical, thus we discard this case. We also reject the value $s=1/2$ which leads to the divergence of $1/(2s-1)$ in Eq.~(\ref{F10}). For $s>1/2$ the rising of parameter $s$ leads to the slower $F_{10}$ decreasing with increasing of $r$ and also to slower increasing of $F_{10}$ when $r\to0$. The behavior of the function $g$ at small $r$ is formed by the second term in Eqs.~(\ref{g1})-(\ref{gs}) related to the charge, whereas at large distances the behavior of the function $g$ is determined by the cosmological term (the first term in Eqs.~(\ref{g1})-(\ref{gs})) leading to asymptotically quadratic increasing of $g$, thereby at large distances the function $g$ approaches asymptotically to the empty anti-de Sitter space and besides for larger values of $s$ this approach is weaker, since the rising of $s$ leads to the more significant contribution of the second term in function $g$ related to the charge of black hole. For certain values of parameters there are zeroes of function $g$ and the largest of them is the event horizon of black hole. Note, that the existence of the event horizon is determined by the parameter $m$ which as mentioned above is related to the black hole mass. The graphs for functions $F_{10}$ and $g$ for various values of $s$ are shown on Fig.~\ref{fig_1}.

The ``2''-component of Eq.~(\ref{em})
\begin{equation}
\left[-2^{s-1}srF_{10}^{2s-1}\left(af+\frac{gF_{12}}{r^2F_{10}}\right)\right]'=0
\end{equation}
with using of Eq.~(\ref{F10}) gives the relation between $F_{10}$ and $F_{12}$
\begin{equation}\label{F12}
\boxed{F_{12}=-\frac{r^2F_{10}}{g}\left(\frac{q_2}{q_1}+af\right),}
\end{equation}
where integration constant $q_2$ we assume proportional to small parameter $a$ ($q_2\sim a$). The combination of ``00'' and ``02'' components of Eqs.~(\ref{grav})-(\ref{stress}) using Eqs.~(\ref{F10}), (\ref{F12}) gives rise to the equation for the non-diagonal metric function $f$
\begin{equation}\label{f}
-rg''f+g'f+rgf''+3gf'=\frac{4q_2F_{10}}{a}.
\end{equation}
The left hand side of the Eq.~(\ref{f}) can be rewritten as $\left[\frac{g^2}{r}\left(\frac{r^2f}{g}\right)'\right]'$ and we obtain
\begin{equation}\label{int}
\boxed{f=\frac{g}{r^2}\left\{\int\frac{r}{g^2}\left[\int\frac{4q_2F_{10}}{a}dr+c_1\right]dr+c_2\right\},}
\end{equation}
where $c_1$ and $c_2$ are integration constants. The inner integral in Eq.~(\ref{int}) can be found in analytic form unlike the outer integral which can not be integrated exactly. Because of this we are going to obtain asymptotic analytic solution for function $f$ for large value of $r$. For this we present the function $g$ in the form $g=-\Lambda r^2(1+z)$, where $z$ is an auxiliary variable which is small at larger distances and its explicit expression is presented in the Appendix. Using the fact that $|z|\ll1$ for large $r$, we replace $1/g^2$ which stands in the outer integrand of Eq.~(\ref{int}) with $1/g^2\approx(\Lambda^2r^4)^{-1}(1-2z)$, namely we neglect all terms higher linear order with respect to the small parameter $z$. In this approximation we obtain the following solutions for function $f$
\begin{equation}\label{f_asympt_1}
f=\frac{g}{r^2}\left\{\frac{1}{r^2}[A_1\ln r+A_2]+\frac{1}{r^4}[A_3\ln^2r+A_4\ln r+A_5]+c_2\right\},\qquad s=1,
\end{equation}
\begin{equation}\label{f_asympt_s}
f=\frac{g}{r^2}\left\{\frac{B_1}{r^2}+\frac{B_2}{r^4}+B_3r^{-\frac{2s}{2s-1}}+B_4r^{-\frac{4s}{2s-1}}+B_5r^{-\frac{2(3s-1)}{2s-1}}+c_2\right\},\qquad s\neq1,
\end{equation}
where constants $A_i$, $B_i$, $i=\overline{1,5}$ are presented in the Appendix. As it is seen from Eqs.~(\ref{f_asympt_1})-(\ref{f_asympt_s}) for large $r$ we have decreasing of function $f$, at that weaker for larger values of $s$. Notice, that for the range $s\in(1/2;1)$ we have $f(r)\to-\Lambda c_2$ when $r\to+\infty$ and in the graphs below we choose the value $c_2=0$, whereas for the integration constants $q_2$ and $c_1$ we choose the general case of nonzero values $q_2$, $c_1$, thereby considering them as the arbitrary parameters of the studied model.

\begin{figure}[h]
\centering
\subfloat{\includegraphics[width=0.35\textwidth]{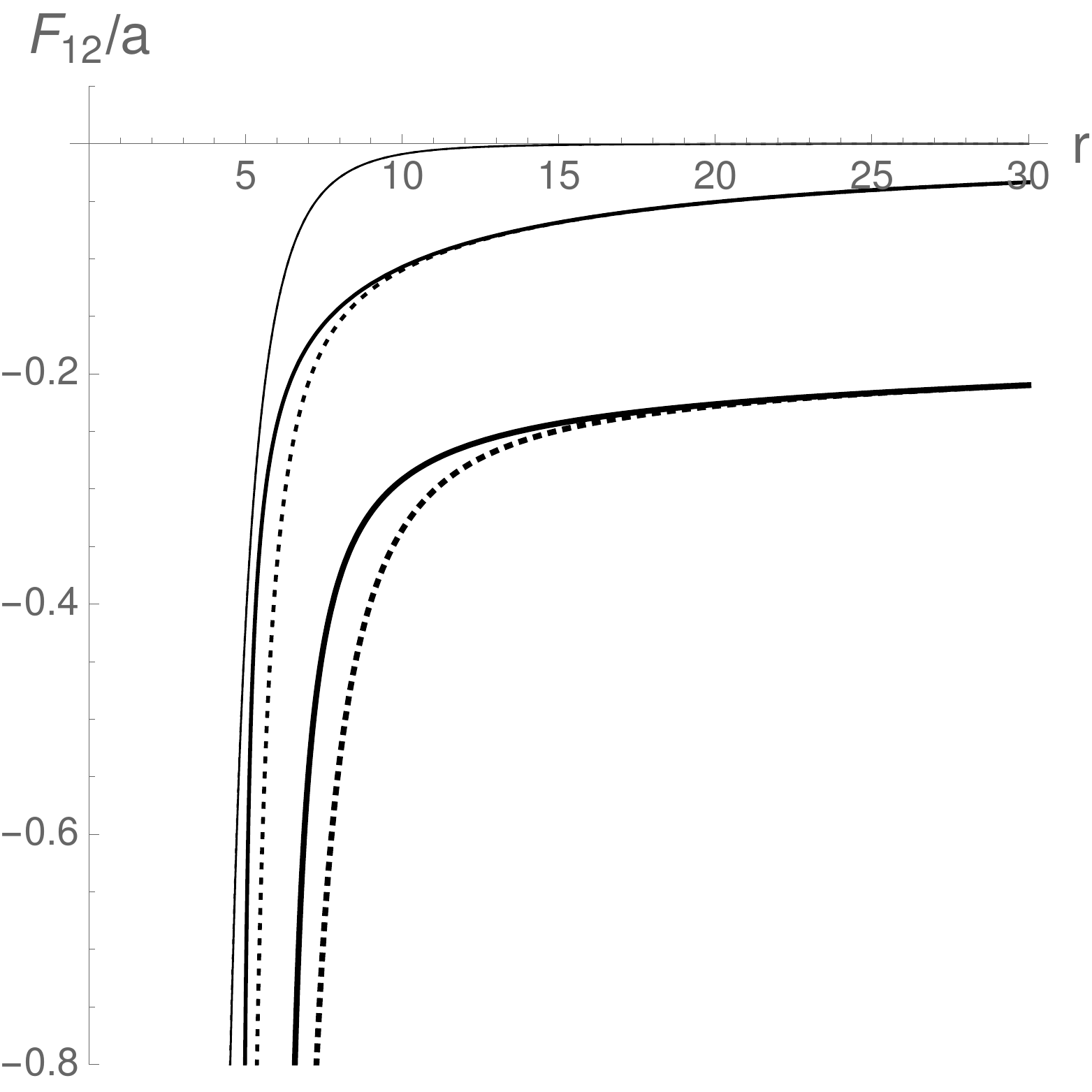}}
\qquad
\subfloat{\includegraphics[width=0.35\textwidth]{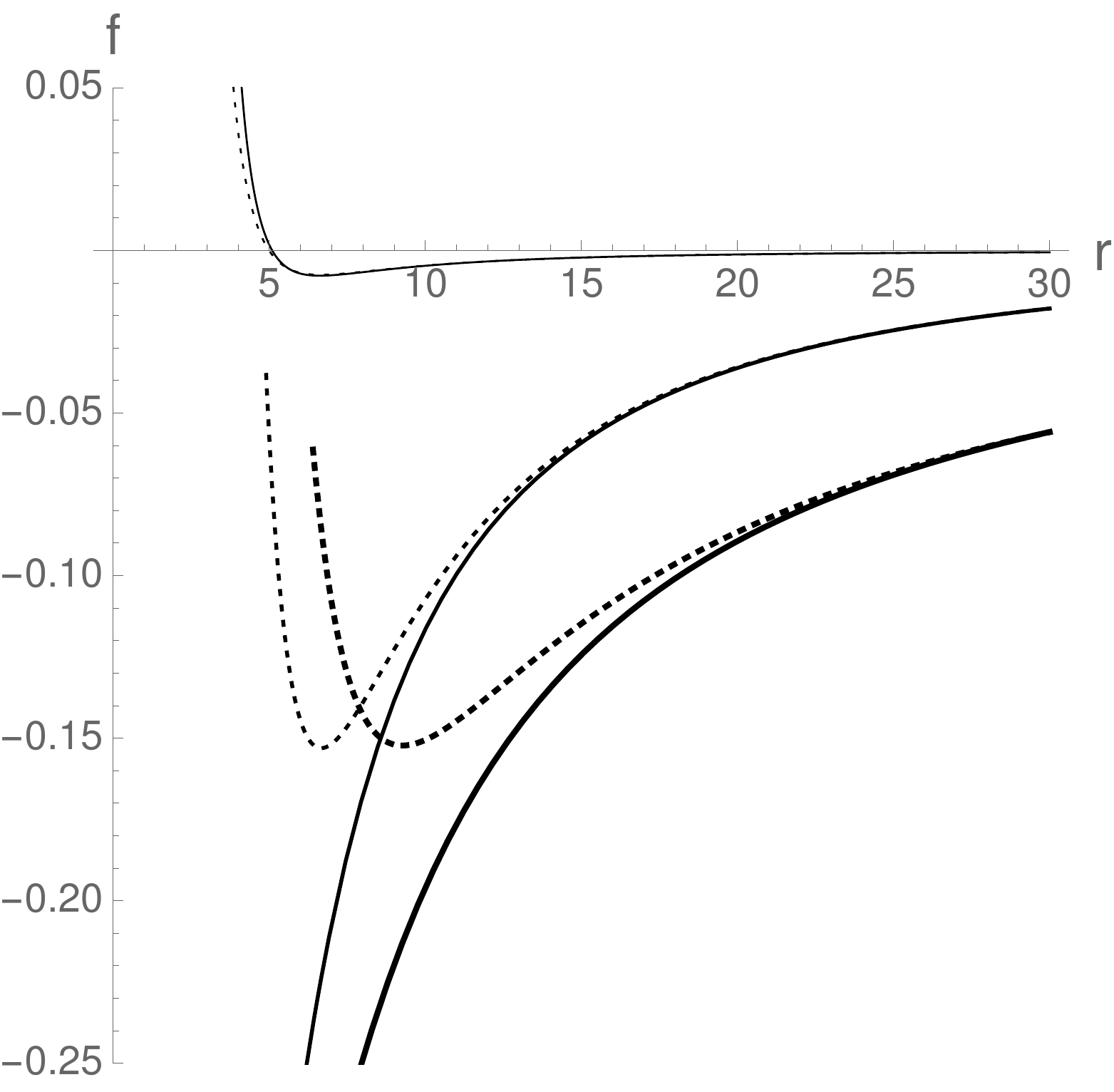}}
\caption{Rescaled magnetic field $F_{12}(r)/a$ (left) and non-diagonal metric function $f(r)$ (right) for $s=0.6$ (thin), $s=1$ (normal), $s=4$ (bold), $q_1=2$, $q_2/a=1$, $m=10$, $c_1=1$, $c_2=0$, $\Lambda=-1$. Dots show the analytic asymptotic solutions, whereas the continuous lines denote the exact solution which is obtained by numerical integration of Eq.~(\ref{int}).}
\label{fig_2}
\end{figure}

Substitution (\ref{f_asympt_1})-(\ref{f_asympt_s}) into Eq.~(\ref{F12}) gives function $F_{12}$ for large $r$. For larger values of power $s$ we have weaker decrease of megnetic field. On Fig.~\ref{fig_2} it is shown the behavior of the analytic asymptotic and also numerical exact solutions for functions $F_{12}$ (precisely $F_{12}/a$) and $f$. This is the first main result of this work.

At the end of this section, note that the, so-called, conformally invariant electromagnetic field mentioned in the Introduction is given for three dimensions by nonlinearity parameter $s=3/4$ and in this case we have traceless stress-energy tensor $\left.T_\mu^\mu\right|_{s=3/4}=0$, whereas the scalar curvature is $\left.R\right|_{s=3/4}=6\Lambda$ and subsequently it vanishes in the absence of the cosmological constant. Besides this, the conformal case $s=3/4$ is included to the range $s\in(1/2;1)$ in which we have a good convergence of the obtained black hole solutions.
\section{Conserved charges}\label{charges}
As it was mentioned above the studied black hole is described by three macroscopic parameters, namely mass, electric charge and angular momentum, the last of which we assume proportional to the small parameter $a$. The electric charge $Q$ can be found using the Gauss's theorem and in the slowly rotating limit we have $Q=q_1/2$ as in the corresponding static case. In order to calculate the mass and angular momentum we use the quasilocal Brown-York formalism \cite{brown93} and the, so-called, counterterm method \cite{kraus99}. For this, besides the bulk action (\ref{act}) one should introduce the boundary terms, namely the Gibbons-Hawking-York (GHY) boundary term and the counterterm related to the presence of the cosmological constant. Then we can write the boundary stress-energy tensor $T_{\mu\nu}^{(b)}$ in the form
\begin{equation}
T_{\mu\nu}^{(b)}=\frac{1}{8\pi}\left(K_{\mu\nu}-h_{\mu\nu}K-h_{\mu\nu}\sqrt{-\Lambda}\right),
\end{equation}
where the first two terms follow from the GHY term whereas the last term corresponds to the counterterm. Here $h_{\mu\nu}=g_{\mu\nu}-n_\mu n_\nu$ is the induced metric, $n_\mu$ is the spacelike unit normal vector, $K_{\mu\nu}=-h_\mu^\delta\nabla_\delta n_\nu$ is the extrinsic curvature and $K$ is the trace of $K_{\mu\nu}$. Then the mass and angular momentum are given by
\begin{equation}
M=\int\limits_0^{2\pi}d\varphi\sqrt{\sigma}T_{\mu\nu}^{(b)}u^\mu\xi^\nu,\qquad r\to+\infty,
\end{equation}
\begin{equation}
J=\int\limits_0^{2\pi}d\varphi\sqrt{\sigma}T_{\mu\nu}^{(b)}u^\mu\zeta^\nu,\qquad r\to+\infty,
\end{equation}
where $u^\mu$ is the timelike unit normal vector, $\xi^\mu=\partial/\partial t$ and $\zeta^\mu=\partial/\partial\varphi$ are Killing vectors. For our case we have $u_\mu=\left(-\sqrt{g},0,0\right)$, $n_\mu=\left(0,1/\sqrt{g},0\right)$, $\xi^\mu=(1,0,0)$, $\zeta^\mu=(0,0,1)$, $\sigma=r^2$ and we obtain
\begin{equation}
M=\left.\frac{1}{4}\left(r\sqrt{-\Lambda g}-g\right)\right|_{r\to+\infty},
\end{equation}
\begin{equation}
J=-\left.\frac{a}{8}\left(r^3f'\right)\right|_{r\to+\infty}.
\end{equation}
For values of pawer $s<1$ we finally calculate
\begin{equation}
M=\frac{m}{8},
\end{equation}
\begin{equation}
J=\frac{a}{8}\left(\frac{c_1}{\Lambda}-2mc_2\right),
\end{equation}
whereas in case $s\geqslant1$ we have divergences due to the behavior of electromagnetic field when $r\to+\infty$ and here one should introduce additional counterterms related to the electromagnetic field to obtain finite conserved charges. Note, that the mass in the small rotation limit is the same as for the corresponding static case.
\section{Thermodynamics}\label{thermodynamics}
\begin{figure}[h]
\centering
\subfloat{\includegraphics[width=0.35\textwidth]{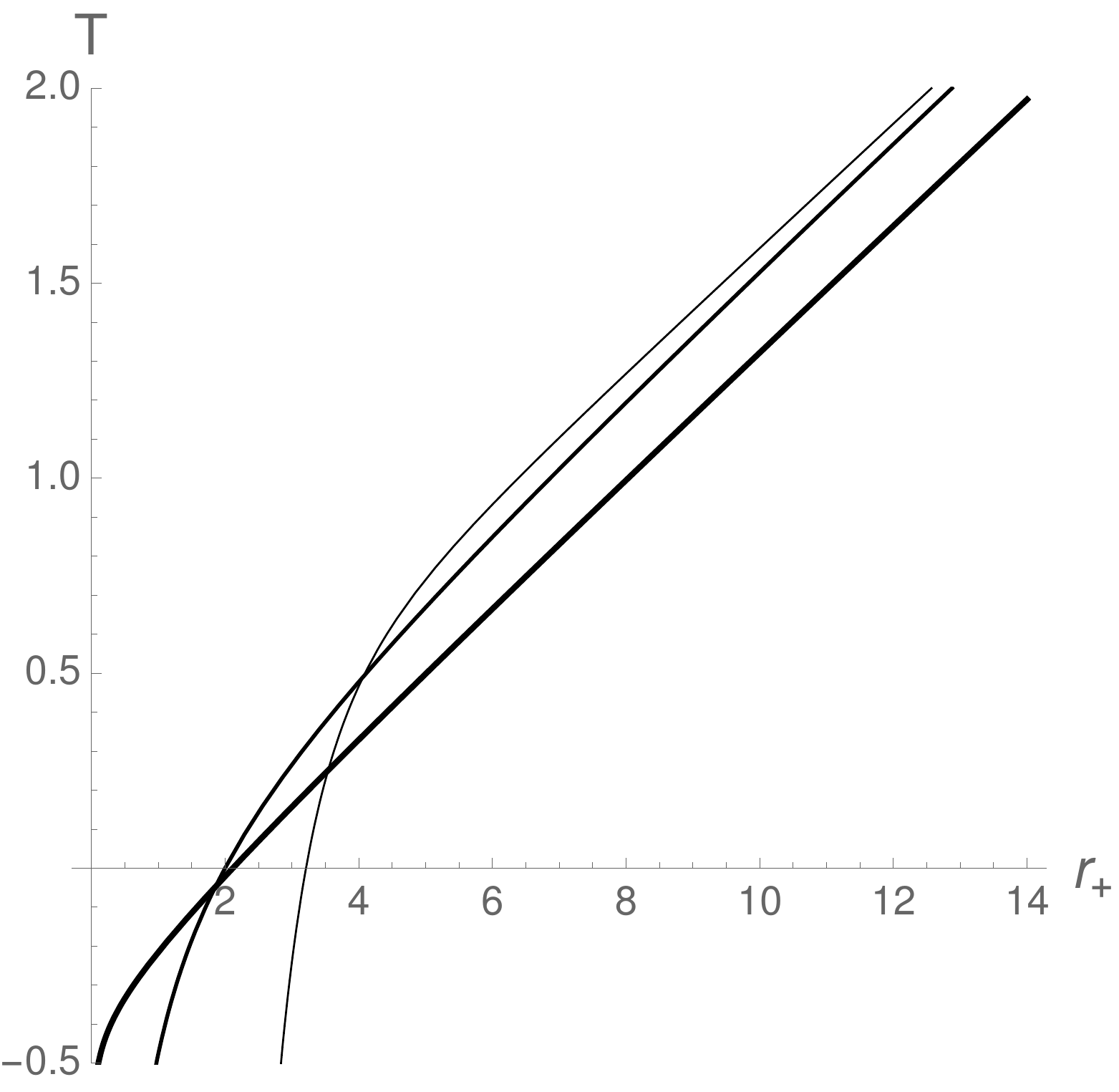}}
\qquad
\subfloat{\includegraphics[width=0.35\textwidth]{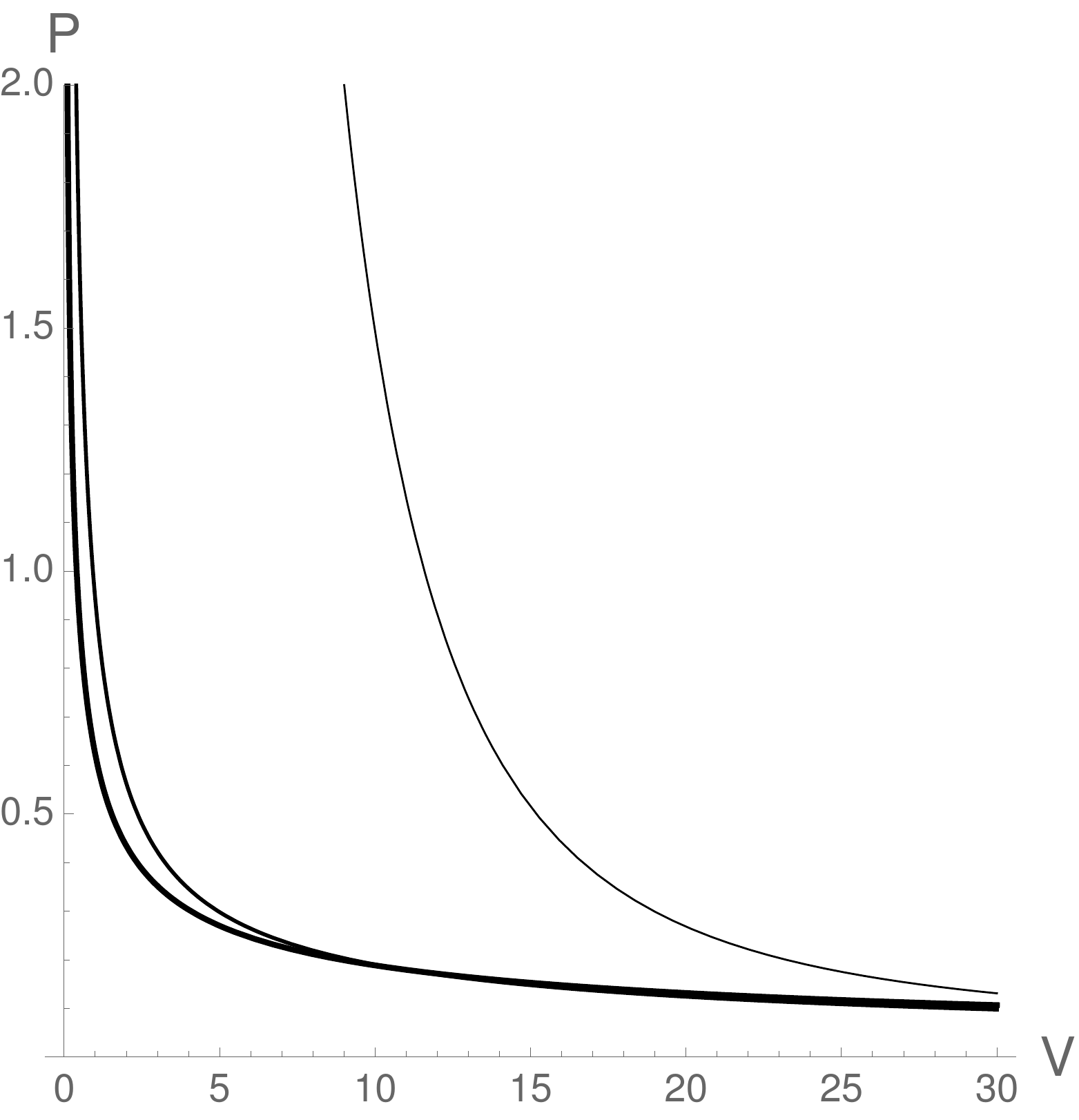}}
\caption{Temperature $T(r_+)$ (left) and isotherms $P(V)$ (right) for $s=0.6$ (thin), $s=1$ (normal) and $s=4$ (bold) for parameters $q_1=2$, $\Lambda=-1$ for left graph and $T=1$, $Q=1$ for right graph.}
\label{fig_3}
\end{figure}

Now we are going to investigate thermodynamic behavior of the black hole solution obtained in the previous Sections. The temperature and the electric potential of the black hole are given, respectively
\begin{equation}
T=\left.\frac{1}{2\pi}\sqrt{-\frac{1}{2}(\nabla_\mu\chi_\nu)(\nabla^\mu\chi^\nu)}\right|_{r=r_+},
\end{equation}
\begin{equation}
U=\left.-A_\mu\chi^\mu\right|_{r=r_+}+C,
\end{equation}
where $r_+$ is the black hole horizon, $\chi^\mu$ is the Killing vector, and since we consider the slowly rotating case and take into account only the terms up to the first order over parameter $a$ we can take the Killing vector in the form similarly as in the static case, namely $\chi^\mu=(1,0,0)$ and likewise the temperature $T=\left.g'/4\pi\right|_{r=r_+}$ and the electric potential $U=\left.-A_0\right|_{r=r_+}+C$ are the same as for the static black hole \cite{tat19}. Here $A_0$ stands for the ``0''-component of electromagnetic potential $A_0=\int F_{10}dr+C$, where $C$ is an integration constant. Note that for $s<1$ the electric potential can be defined with respect to infinity and in this case we have $C=0$. Accordingly, $T$ and $U$ are written in the following form
\begin{equation}\label{T}
T=-\frac{1}{2\pi}\left[\Lambda r_++2^{s-1}(2s-1)\left(\frac{q_1}{2^{s-1}s}\right)^\frac{2s}{2s-1}r_+^{-\frac{1}{2s-1}}\right],
\end{equation}
\begin{equation}
U=-q_1\ln r_+,\qquad s=1,
\end{equation}
\begin{equation}\label{Us}
U=-\frac{2s-1}{2(s-1)}\left(\frac{q_1}{2^{s-1}s}\right)^\frac{1}{2s-1}r_+^\frac{2(s-1)}{2s-1},\qquad s\neq1.
\end{equation}
A graph for temperature is shown at the left hand side of the Fig.~\ref{fig_3} and which demonstrates that temperature is monotonous increasing function of black hole horizon.

As it was obtained above the thermodynamic mass function $M$ also takes the form, the same as for static case
\begin{equation}
M=\frac{m(r_+)}{8}.
\end{equation}
Taking into account the fact that the small rotation does not affect the area of the event horizon of black hole \cite{hen10} and using the entropy area law, namely the fact that entropy $S=A/4$ is a quarter of horizon's area $A$ one can obtain a relation for black hole entropy $S$ in the small rotation limit as $S=\pi r_+/2$ which coincides with static case. Besides, using the obtained above relation for the total electric charge $Q$ we write the first law of black hole thermodynamics, which also is the same as in the static case.
\begin{equation}
dM=TdS+UdQ,
\end{equation}
where $T$ and $U$ are given by Eqs.~(\ref{T})-(\ref{Us}).

Finally we are going to extend the thermodynamic space considering the cosmological constant $\Lambda$ as one of the thermodynamic quantities, namely the thermodynamic pressure \cite{kub12}
\begin{equation}
P=-\frac{\Lambda}{8\pi}.
\end{equation}
Introducing the conjugate thermodynamic volume, the first law in the extended phase space thermodynamics can be written with the pressure-volume term
\begin{equation}
dM=TdS+UdQ+VdP,
\end{equation}
where the thermodynamic function $M$ now can be identified with enthalpy, whereas the volume is equal $V=\pi r_+^2=4S^2/\pi$. Using Eq.~(\ref{T}) for the temperature and relations with $r_+$, $q_1$, $\Lambda$ and $V$, $Q$, $P$, respectively, we write the equation of state $P(V,T,Q)$ in the form
\begin{equation}\label{P}
P=\frac{\sqrt{\pi}T}{4\sqrt{V}}+2^{s-4}\pi^{-\frac{s-1}{2s-1}}(2s-1)\left(\frac{2Q}{2^{s-1}s}\right)^\frac{2s}{2s-1}V^{-\frac{s}{2s-1}}.
\end{equation}
It is known that the equation of state allows to check that black holes might have some phase transitions \cite{kub12}. But in our case the system does not have the phase transition, one can easily check this fact using the Eq.~(\ref{P}) or looking at the right graph of the Fig.~\ref{fig_3} where the isotherms demonstrate monotonous behavior.

The heat capacity can be calculated as
\begin{equation}
C=T\left(\frac{\partial S}{\partial T}\right)_{Q,P}.
\end{equation}
Using the Eq.~(\ref{T}) and relations between $r_+$, $q_1$, $\Lambda$ and $S$, $Q$, $P$, respectively one can obtain the heat capacity
\begin{equation}
C=\pi\frac{\Lambda r_++2^{s-1}(2s-1)\left(\frac{q_1}{2^{s-1}s}\right)^\frac{2s}{2s-1}r_+^{-\frac{1}{2s-1}}}{2\Lambda-2^s\left(\frac{q_1}{2^{s-1}s}\right)^\frac{2s}{2s-1}r_+^{-\frac{2s}{2s-1}}}
\end{equation}
which coincides with the heat capacity of the static black hole and it was disscussed in our previous paper \cite{tat19}.

Thus all considered thermodynamics including the extended thermodynamic space is completely identical to the static case \cite{tat19}, that is the small rotation of black hole does not affect on its thermodynamics as it takes place in higher dimensions \cite{hen10}. That is the second main result of this work, although in some sense it is expected.
\section{Conclusions}\label{conclusions}
In this paper we have obtained the three-dimensional slowly rotating black hole solution with the power Maxwell nonlinear electromagnetic field in the presence of the cosmological constant. Obtained solution is described by the two metric functions, namely diagonal function $g(r)$ and non-diagonal function $f(r)$ and also two components of electromagnetic field tensor, namely the radial electric field $F_{10}(r)$ and the magnetic field $F_{12}(r)$. It was found that in the small rotation limit, which is defined by small rotation parameter $a$ related to the black hole angular momentum, the functions $g(r)$, $F_{10}(r)$ and the investigated thermodynamics, including the consideration of the extended phase space thermodynamics, are the same as for corresponding static case.

The black hole temperature and isotherms have monotonous behavior without of a phase transition. The mass, entropy and electric potential satisfy the first law of black hole thermodynamics and which does not include a term related to the angular momentum of rotating black hole. Such identity behavior of thermodynamics of slowly rotating black hole with respect to the static case is due to the fact that the rotation effects influenced on this appear in higher orders over parameter $a$ and thus are absent in the small rotation limit.

Simultaneously, the presence of small rotation is described by the functions $f(r)$ and $F_{12}(r)$ which also give the small contribution, since $F_{12}(r)$ is proportional to the small parameter $a$ whereas the function $f(r)$ enters to the metric (\ref{met}) as $af(r)$. Therefore we see that the key features caused by the small rotation of three-dimension black hole with the power Maxwell field are similar to those for higher-dimensional slowly rotating black holes in the Einstein-power-Maxwell theory and which once again confirms the fact that a three-dimensional black holes possess the main properties which have their higher-dimensional analogues.
\section*{Acknowledgments}
This work was partly supported by the projects $\Phi\Phi$-83$\Phi$ (No. 0119U002203), $\Phi\Phi$-63Hp (No. 0117U007190) from the Ministry of Education and Science of Ukraine.
\section*{Appendix}
$\displaystyle z=\frac{1}{\Lambda r^2}\left(2q_1^2\ln r+m\right),\qquad s=1$,\\\\
$\displaystyle z=\frac{1}{\Lambda r^2}\left(\frac{2^{s-1}(2s-1)^2}{s-1}\left(\frac{q_1}{2^{s-1}s}\right)^\frac{2s}{2s-1}r^\frac{2(s-1)}{2s-1}+m\right),\qquad s\neq1$,\\\\
$\displaystyle A_1=-\frac{2q_1q_2}{a\Lambda^2}$,\\\\
$\displaystyle A_2=-\frac{2q_1q_2+ac_1}{2a\Lambda^2}$,\\\\
$\displaystyle A_3=\frac{4q_1^3q_2}{a\Lambda^3}$,\\\\
$\displaystyle A_4=\frac{2q_1^3q_2-2q_1q_2m+aq_1^2c_1}{a\Lambda^3}$,\\\\
$\displaystyle A_5=\frac{2q_1^3q_2-2q_1q_2m-2amc_1+aq_1^2c_1}{4a\Lambda^3}$,\\\\
$\displaystyle B_1=-\frac{c_1}{2\Lambda^2}$,\\\\
$\displaystyle B_2=-\frac{mc_1}{2\Lambda^3}$,\\\\
$\displaystyle B_3=-\frac{(2s-1)^2q_2}{as(s-1)\Lambda^2}\left(\frac{q_1}{2^{s-1}s}\right)^\frac{1}{2s-1}$,\\\\
$\displaystyle B_4=\frac{2^{s-1}(2s-1)^4q_2}{as(s-1)^2\Lambda^3}\left(\frac{q_1}{2^{s-1}s}\right)^\frac{2s+1}{2s-1}$,\\\\
$\displaystyle B_5=\frac{(2s-1)^2}{(s-1)(3s-1)\Lambda^3}\left[\frac{(2s-1)q_1c_1}{s}-\frac{2q_2m}{a}\right]\left(\frac{q_1}{2^{s-1}s}\right)^\frac{1}{2s-1}$.

\end{document}